\documentclass[useAMS,usenatbib]{mn2e}
\usepackage{graphicx}
\usepackage{amsmath}
\usepackage{amssymb}

\newcommand{\be}{\begin{equation}}                                 
\newcommand{\ee}{\end{equation}}                                   
\newcommand{\bea}{\begin{eqnarray}}                                
\newcommand{\eea}{\end{eqnarray}}

\title[Stable, levitating, optically thin atmospheres]{Stable, levitating, optically thin atmospheres of Eddington-luminosity neutron stars}
\author[Wielgus, Klu\'zniak, S\c{a}dowski, Narayan \& Abramowicz]
{ {M. Wielgus$^{1,2}$\thanks{E-mail:maciek.wielgus@gmail.com}, W. Klu\'zniak$^{1}$\thanks{E-mail:wlodek@camk.edu.pl}, {A. S\c{a}dowski}$^{3,4}$\thanks{E-mail:asadowski@mit.edu}, {R. Narayan}$^{4}$ and {M. Abramowicz}$^{1,5}$ 
}\\
$^{1}$ Copernicus Astronomical Center,
 ul. Bartycka 18, PL 00-716 Warszawa, Poland\\
$^{2}$ Institute of Micromechanics and Photonics, Warsaw University of Technology, ul. \'Sw. A. Boboli 8, 02-525, Warszawa,
Poland\\
$^{3}$ MIT Kavli Institute for Astrophysics and Space Research 77
Massachusetts Ave, Cambridge, MA 02139, USA\\
$^{4}$ Harvard-Smithsonian Center for Astrophysics, 60 Garden Street, Cambridge, MA 02138, USA\\
$^{5}$Physics Department, Gothenburg University, 412-96 G\"oteborg, Sweden}
\begin{document}

\date{Accepted ***. Received **; in original form 2015 May 9}

\pagerange{\pageref{firstpage}--\pageref{lastpage}} \pubyear{2015}

\maketitle

\label{firstpage}

\begin{abstract}
  In general relativity static  gaseous atmospheres may
    be in hydrostatic balance in the absence of a supporting stellar
    surface, provided that the luminosity is close to the Eddington
    value.  We construct analytic models of optically thin,
    spherically symmetric shells supported by the radiation pressure
    of a luminous central body in the Schwarzschild metric. {Opacity is assumed to be dominated by Thomson scattering. The inner parts of the atmospheres, where the luminosity locally has supercritical values, are characterized by a density and pressure inversion. The atmospheres are convectively and Rayleigh-Taylor stable, and there is no outflow of gas.}
\end{abstract}

%
   \section{Introduction}
Several systems with super-Eddington luminosity have been
reported \citep[e.g.,][]{2006csxs.book..157M}, the ``LMC transient'' A0535-668
\citep{1983ARA&A..21...13B} being oft discussed.
Recently, interest in very luminous neutron stars has been revived
with the discovery of a clear example of a neutron star with super-Eddington
flux in the guise of
the ultraluminous $1.37\,$s pulsar NuSTAR J095551+6940.8
in the nearby galaxy M82 \citep{Bachetti2014}.
 It is clear that for some accreting neutron
stars, radiation pressure may exceed the pull of gravity close to
their surface (at least in a certain solid angle if the radiation is beamed).

In this paper we report the presence of a new type of atmospheric
solutions in general relativity (GR) for neutron stars radiating at
nearly Eddington luminosities.  These solutions are qualitatively
different from the ones obtained in Newtonian physics. 
In the classical solutions the atmospheric density increases
monotonically as the stellar surface is approached, and this remains
true in previously obtained atmospheric solutions in GR
\citep[e.g.,][]{Paczynski1986}, where the atmospheres in question have
always been supported at their base by the stellar surface.  Here, we
report solutions in which the atmosphere has the form of a shell
suspended above the stellar surface. The maximum of pressure is
attained on a surface enclosing a volume in which the star is
contained (together with some space around it), and the atmospheric
density and pressure drop precipitously on both sides of that surface,
i.e.,  in the radial direction away from the star (as is usually the
case), but also in the radial direction towards the star.

This paper discusses the simplest case of optically thin, spherically
symmetric atmospheres in the Schwarzschild metric,
which admit of analytic solutions. Optically thick atmospheres
require numerical treatment of radiative transfer, and these will
be reported elsewhere \citep{Wielgus2015}.

We consider atmospheres consisting of pure ionized hydrogen.
All the results are given for the Schwarzschild spacetime.
The luminosity of the star will be parametrized by 
the ratio of the luminosity observed at infinity
to the Eddington luminosity,
\be
\lambda=L_\infty/L_{\text{Edd}},
\label{lambda}
\ee
with the standard expression for the latter,
$L_{\text{Edd}} = {4 \pi GM m_pc }/{\sigma_{\text T}}$.
 The stellar radius will be denoted by $R_*$, it is taken to correspond
to a canonical neutron star, e.g., $R_*\approx5GM/c^2$,
with $M\approx 1.4M_\odot$ being the stellar mass.

%
\section{Balance between gravity and radiation pressure}
\label{s:thin}

In the Schwarzschild metric,
the stellar luminosity (which we take to be constant in time), as measured by a local observer at radius $r$,
declines with the radial coordinate distance $r$ as
\be
L(r)=L_\infty\left(1-\frac{2GM}{rc^2}\right)^{-1},
\label{luminosity}
\ee
%
and a static balance between gravity and radiation pressure
owing to Thomson scattering can only be achieved at one radius,
at which
\be
L(r)=L_{\text{Edd}}(1- R_{\text{S}}/r)^{-1/2},
\label{localEdd}
\ee
where $R_{\text{S}}={2GM}/{rc^2}$ is the Schwarzschild radius
\citep[e.g., ][]{Phinney1987,Stahl2013}.
That radius can be written as
\be
r= R_{\text{S}}/\left(1-\lambda^2\right) \equiv R_{\text{ECS}}.
\label{ECS}
\ee
{The sphere at $R_{\text{ECS}}$ has been called the Eddington capture sphere (ECS) by \citet{Stahl2012}}.
These results have been derived rigorously within the mathematical formalism
of Einstein's general relativity, and they can be intuitively
understood in terms of the different dependence
on redshift of the luminosity and of the ``gravitational'' acceleration.
 Eq.~(\ref{luminosity}) reflects the dependence of luminosity
at infinity on two redshift  factors $(1+z)^{-1}$,
one for the energy of the photons, and one for the rate of their arrival.
However, the acceleration of a static observer in the Schwarzschild metric
scales with only one factor of $(1+z)$, where the redshift is defined by
\be
1+ z(r) =(1- R_{\text{S}}/r)^{-1/2}.
\label{red}
\ee
Hence, the condition for local
balance between acceleration and the momentum flux of photons
(multiplied by the Thomson opacity) is given by Eq.~(\ref{localEdd}),
and contains one factor of  $(1+z)$, which introduces
into  the balance condition a dependence on the radial coordinate.
 This is in contrast with Newtonian
gravity (which can formally be recovered in the limit
 $R_{\text{S}}\rightarrow 0$, or $z\rightarrow 0$),
where the balance condition has no radial dependence,
and therefore is either satisfied, or not satisfied,
equally at all distances from the source.

Numerous authors
have shown that test particles initially orbiting the star (at various
radii) may settle on the spherical surface at $r=R_{\text{ECS}}>R_*$,
provided that
\be
(1- R_{\text{S}}/R_*)^{1/2}<\lambda<1,
\label{range}
\ee
%
their angular momentum having been removed by radiation drag
\citep{Bini2009,Oh2011,Stahl2012,Mishra2014}.
In fact, any point on the ECS is a position of {\sl stable} equilibrium
 in the radial direction \citep{Abramowicz1990},
and neutral equilibrium in directions tangent
 to the ECS surface \citep{Stahl2012}. 
Note that the redshift $z(r)$
attains the following value on the ECS,
\be
z_{\text{ECS}}\equiv z(R_{\text{ECS}})=1/\lambda-1.
\label{zecs}
\ee
%

We now take the step of replacing test particles on the ECS with a fluid.
{\bf For simplicity, let us assume\footnote{The optically thick case will be reated in a separate paper \citep{Wielgus2015}.}} that the fluid is optically thin. This will
allow us to decouple the temperature of the fluid from the radiation
field and to find analytic solutions of the atmospheric structure. 
It is clear that the condition of hydrostatic
equilibrium will be achieved if the pressure gradient is in the radial
direction and its magnitude compensates the imbalance between
radiation pressure and gravity. In particular, the pressure must reach
a maximum value at  $r=R_{\text{ECS}}$, and must decrease in both
directions, towards and away from the star, as the distance to the ECS
 grows and so does with it the imbalance between radiation
pressure and gravity.

For atmospheric temperatures of no more than a few keV, we can safely
assume  that the energy density of the fluid is given
by its baryon rest mass energy density $\rho c^2$. In other words,
one can neglect the contribution of pressure $p$ and the internal energy
$\epsilon$ to the energy density, $\rho c^2\gg p +\epsilon$.
The equation of hydrostatic equilibrium then becomes
\begin{equation}
\frac{1}{\rho} \frac{\text{d} p}{\text{d} r} = 
-\frac{GM}{r^2\left(1-{R_{\text{S}}}/{r}\right)}\left[1 -
{\lambda}{\left(1-\frac{R_{\text{S}}}{r}\right)^{-1/2}}\right] \ \text{,}
\label{equilibrium}
\end{equation}
where the last term reflects the radiation pressure owing to
Thomson scattering, with the
luminosity of the star scaled by the Eddington luminosity
per Eq.~(\ref{lambda}). In the limit ${R_{\text{S}}}/r\rightarrow0$
one recovers the classical Eddington balance at $\lambda=1$, while
in general the pressure gradient vanishes for $r=R_{\text{ECS}}$,
cf. Eq.~(\ref{ECS}).

Eq.~(\ref{equilibrium}) is readily solved for a polytropic
or an isothermal atmosphere, as it can be integrated over the redshift
to yield
\be
\int \frac{\text{d} p}{\rho c^2} = \ln (1 + z) - \lambda  z .
\label{ilibrium}
\ee
For an ideal gas equation of state, ${p}={\rho} k_B T/(\mu m_p)$
with $k_B$ the Boltzmann constant, $m_p$ the proton mass, 
and $\mu{\bf =1/2}$ the mean molecular weight,
we obtain the following analytic solutions.
%
\begin{figure}
\includegraphics[width = 3.5in,trim = 0mm 5mm 0mm 5mm]{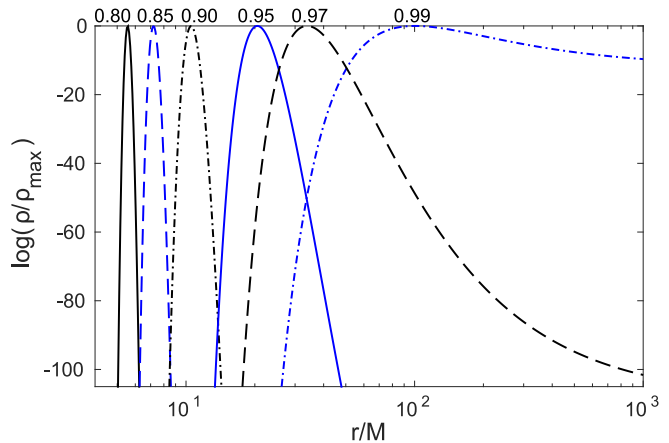} 
\caption{Density profiles for the isothermal solutions with $T=10^7\,$K.
 Each curve corresponds to a different value of the luminosity parameter:
 $\lambda=0.8$, 0.85, 0.9, 0.95, 0.97, 0.99 from left to right (as labeled at the maxima).}
\label{fig:pnewtonian}
\end{figure}

\section{Isothermal atmosphere}
\label{sub:isothermal}
For an isothermal atmosphere at temperature $T$, the integral in
Eq.~(\ref{ilibrium}) yields
\be
\frac{k_BT}{\mu m_pc^2} \ln\left(\frac{\rho}{\rho_0}\right)
=\ln (1 + z) - \lambda  z.
\label{isothermal1} 
\ee
with $z$ defined in Eq.~(\ref{red}).
In the GR case, the maximum value of density is attained at the ECS,
assuming a luminosity in the range given by Eq.~(\ref{range}), and the
density can be taken to be proportional to this maximum value
$\rho_{\text{max}}=\rho(z_{\text{ECS}})$,
\bea
\rho(z) =& \rho_{\text{max}}\times [\lambda (1 + z)]^{(\mu m_pc^2)/(k_BT)}\nonumber\\
& \times \exp\left\{[ 1- \lambda (1 + z)] 
{(\mu m_pc^2)/(k_BT)}\right\}.
\label{isothermal2} 
\eea
Thus, as expected, the density decreases in both directions away from
the ECS, giving the atmosphere a shell-like form.
The same result, given as a function of the radius, is
\bea
\rho(r) =& \rho_{\text{max}}\times [\lambda(1-R_{\text{S}}/r)^{-1/2}]^{(\mu m_pc^2)/(k_BT)}
\label{isothermal3} 
\\
&\times  \exp\left\{[ 1- \lambda (1-R_{\text{S}}/r)^{-1/2}] 
{(\mu m_pc^2)/(k_BT)}\right\}.\nonumber
\eea
However, far from the star this solution is unphysical, as the
mass in an isothermal atmosphere would be infinite---note that at $z=0$,
 i.e., as $r\rightarrow\infty$,
the density tends to a finite value,
$\rho\rightarrow \rho_\infty\equiv\rho_{\text{max}}\times
 [\lambda \exp{(1- \lambda)}]^{(\mu m_pc^2)/(k_BT)}$. 

The Newtonian Eddington limit 
is recovered with a different normalization from the one in
Eqs.~(\ref{isothermal2})-(\ref{isothermal3}).
To first order in $z$, from any of the three 
Eqs.~(\ref{isothermal1})-(\ref{isothermal3}) one gets,
$\rho(r)=\rho_\infty\times\exp\left[{\mu m_pGM(1-\lambda)}/{(k_BTr)}\right]$.
The classical Eddington limit, $dp/dr=0$, or
$\rho=\text{const}$ for an isothermal atmosphere,
is now obtained for $\lambda =1$, while the
Newtonian limit of an exponentially decaying atmosphere
holds for $|r-r_0|\ll r_0$
\be
\rho(r) \approx
\rho_0 \exp\left[-\frac{\mu m_p}{k_BT}\frac{GM(1-\lambda)}{r_0^2}(r-r_0)\right],
\label{newtonian}
\ee
with arbitrary constants $\rho_0$ and $r_0$ of dimension density and length,
respectively.

Although we have assumed Thomson scattering, this is not strictly necessary.
In fact, all the above considerations in Section~\ref{sub:isothermal},
as well as Eq.~(\ref{ECS}), remain valid if we replace the Thomson value,
 $\sigma_{\text T}$,
of the photon electron scattering cross section by a more general cross-section
 $\sigma(T)$, e.g., the Klein-Nishina formula,
provided that we make the substitution 
$\lambda\rightarrow\lambda_1=\lambda a_1$, with $a_1= \sigma(T)/\sigma_{\text T}$.

 We stress once again that unlike the decaying atmosphere of the Newtonian
limit in Eq.~(\ref{newtonian}),
the GR solution in 
Eqs.~(\ref{isothermal1})-(\ref{isothermal3}) is not monotonically decreasing
with radius everywhere, but instead is  monotonically increasing
 for $r<R_{\text{ECS}}$
up to the maximum on the ECS, and is only monotonically decreasing
for $r>R_{\text{ECS}}$, as illustrated in Fig.~(\ref{fig:pnewtonian}). 

%
\section{Polytropic atmosphere}

For a polytrope, $p=K\rho^\Gamma$, the specific enthalpy 
$w =  ({p}/{\rho}) {\Gamma}/({\Gamma-1})$ following from
Eq.~(\ref{equilibrium}) or Eq.~(\ref{ilibrium})  is
\be
w(z)=w_{\text{ECS}}
 + c^2\left\{\ln \left[\lambda (1 + z) \right] - \lambda (1 + z) +1\right\}
\label{enth}
\ee
and it attains its maximum value, 
$w_{\text{ECS}}=w(z_{\text{ECS}})$, at $r=R_{\text{ECS}}$, see Eq.~(\ref{zecs}).

\begin{figure}
\includegraphics[width = 3.5in,trim = 0mm 5mm 0mm 5mm]{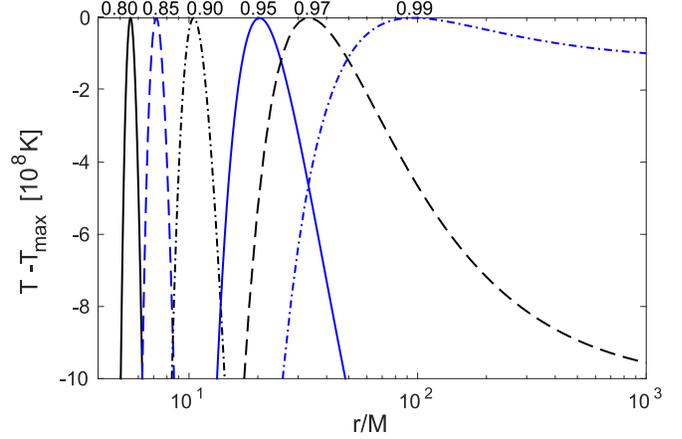} 
\caption{Temperature profiles for the polytropic solutions. 
 Each curve corresponds to a different value of the luminosity parameter:
 $\lambda=0.8$, 0.85, 0.9, 0.95, 0.97, 0.99 from left to right.
Note that the temperature difference on the $y$ axis is in units of $10^8\,$K.}
\label{fig:Tpolytrope}
\end{figure}

Making use of the ideal gas equation of state  we obtain
 a corresponding solution for the temperature
\bea
\label{Tpoly}
T(r) -  T_{\text{max}} = & \\
 - \left(\frac{\Gamma-1}{\Gamma }\frac{\mu m_pc^2}{k_B}\right) & 
\left\{ \frac{\lambda}{(1-{R_{\text{S}}}/{r})^{1/2}}
 - \ln \left[ \frac{\lambda}{(1-R_{\text{S}}/r)^{1/2}} \right] - 1 \right\},
 \nonumber
\eea
also attaining its maximum value on the ECS.
Note that for any given value of the luminosity, $\lambda$,
the temperature falls of on both sides of the ECS in a universal
way (Fig.~\ref{fig:Tpolytrope}) independently of the value of the 
maximum temperature $T_{\text{max}}$, which is just an additive constant
in Eq.~(\ref{Tpoly}). Of course the temperature cannot go to negative values, so the plot must be cut off at $T-T_{\text{max}}=-T_{\text{max}}$. Fig.~\ref{fig:Tpolytrope} corresponds to $T_{\text{max}}=10^9\,$K, i.e., a hot $\sim 100\,$keV corona.

The density scales with its arbitrary value at the ECS,
 $\rho_{\text{max}}\equiv\rho(R_{\text{ECS}})$,
which is proportional to the total mass of the atmosphere:
\be
\rho(r) = \rho_0\left[\ln(1-R_{\text{S}}/r)^{-1/2} 
 -\lambda(1-R_{\text{S}}/r)^{-1/2} +\lambda \right]^\frac{1}{\Gamma-1},
\ee
with $\rho(R_{\text{ECS}})= \rho_0 [\lambda-\ln \lambda -1]^{1/(\Gamma-1)}$.

\begin{figure}
\includegraphics[width = 3.5 in,trim = 0mm 5mm -5mm 5mm]{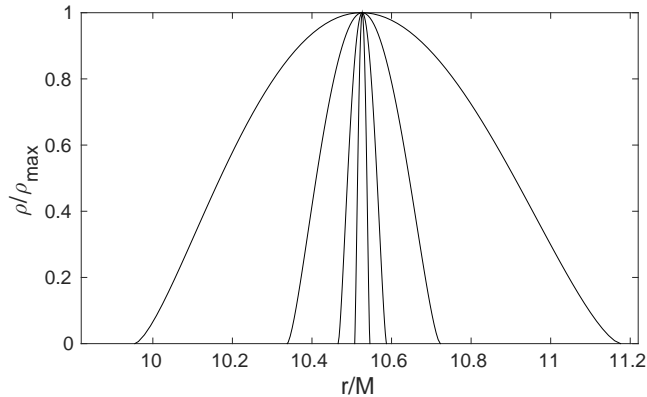}
\caption{Density profiles of polytropic
Thomson-scattering atmospheres for $\lambda = 0.9$.
For a luminosity this large, the atmosphere is well separated from the
neutron star surface (at $R_*/M\approx 5$).
Temperatures from outside in (more extended atmospheres to less
extended atmospheres) are $T_{\text{max}}=5\cdot10^7 \,$K,
$5\cdot10^6 \,$K, $5\cdot10^5 \,$K, $5\cdot10^4 \,$K, respectively,
with $\mu=1/2$ and $\Gamma=5/3$ in all cases.
The density maxima are at $R_{\text{ECS}} = 10.5 \, GM/c^2$.}
\label{fig:Polydensity}
\end{figure}

This being a polytrope, the atmosphere has a sharp edge,
the enthalpy, temperature and density vanishing at the top of the atmosphere.
In fact, when the condition of Eq.~(\ref{range}) is satisfied,
there are two ``tops'' of the atmosphere (Fig.~\ref{fig:Polydensity}),
 one  at $r_+>R_{\text{ECS}}$,
with $z_+\equiv z(r_+)<z_{\text{ECS}}$, and one  at $r_-<R_{\text{ECS}}$,
with  $z_-\equiv z(r_-)>z_{\text{ECS}}$. The redshift at the top of the atmosphere
is given by the condition
\be
\left\{\ln \left[\lambda (1 + z_\pm) \right] - \lambda (1 + z_\pm) +1\right\}
=-w_{\text{ECS}}/ c^2 .
\label{entha}
\ee
 For maximum temperatures not much larger than a few keV, the
atmospheric profile is quite symmetric with respect to the sign of
$r-R_{\text{ECS}}$ (Fig.~\ref{fig:Polydensity}), and the height of the
atmosphere is proportional to the speed of sound at its ``base,''
$c_s=\sqrt{(\Gamma-1)w_{\text{ECS}}}$.

Indeed, for $H\equiv |r_\pm-R_{\text{ECS}}|\ll R_{\text{ECS}}$,
\be
\frac{H}{R_{\text{ECS}}}={\frac{\sqrt2\lambda^2}{\sqrt{\Gamma-1}}}
\frac{c_sc}{v_{\text{K}}^2},
\ee
where $v_{\text{K}}=\sqrt{GM/R_{\text{ECS}}}$ is the Keplerian orbital 
velocity at the ECS.
This is reminiscent of the result for accretion disks,
where $H/r\approx c_s/v_{\text{K}}$ \citep{SS73}, however the ECS atmospheres
are more extended, $H$ here being enhanced by a factor of $c/v_{\text{K}}$.
In terms of the temperature at the base of the atmospheric shell, this reads
\be
\frac{H}{R_{\text{ECS}}}=
\frac{2\lambda^2}{1-\lambda^2}\left({\frac{2\Gamma}{\Gamma-1}}\right)^{1/2}
\left(\frac{k_BT_{\text{max}}}{\mu m_pc^2}\right)^{1/2}.
\ee
For $k_BT_{\text{max}}=2\,$keV, and $L_\infty\approx L_{\text{Edd}}$
the shell thickness works out to be
 $H\approx 2\cdot 10^{-3}(1-\lambda)^{-2}R_{\text{S}}$.
 For a neutron star this gives 
 $H\approx 10(1-\lambda)^{-2}{\text{m}}$, i.e., 
$H=1\,$km for $\lambda=0.9$, and $H=4\,$km for  $\lambda=0.95$.
One can see that as the luminosity increases, the thickness of the
levitating atmospheric shell increases more quickly than the radius
of the ECS.

\begin{figure}
\includegraphics[width = 3.5 in,trim = 0mm 5mm -5mm 5mm]{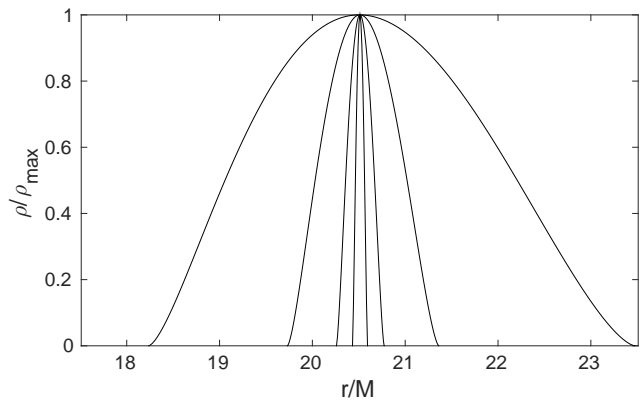}  
\caption{Same as Fig.~\ref{fig:Polydensity}, but for $\lambda = 0.95$.
The density maxima are at $R_{\text{ECS}} = 20.5 \, GM/c^2$.}
\label{fig:Polydensitya}
\end{figure}
%
\section{Discussion}
We have shown that luminous compact stars (e.g., neutron stars) in Einstein's general relativity
may have atmospheres which are detached from the surface. {This may have applications to astrophysical phenomena of those compact objects in which the luminosity is close to its Eddington value, either on a quasi-permanent basis or in a transient state. The former case includes highly luminous neutron stars accreting matter from low-mass stellar companions, such as the Z-sources (which include the brightest X-ray source in the sky, Sco X-1).
One would expect quasi-spherical coronae in such sources to be described by our solutions with implications for the hard X-ray spectra in these sources, and possibly also for their quasiperiodic modulations. The latter include} X-ray bursts, some of which
exhibit the phenomenon of radius expansion \citep[e.g.,][]{2006csxs.book..113S}.
However, here we have only considered optically thin atmospheric shells,
whereas the atmospheres of X-ray bursters are optically thick. For this reason we are postponing a detailed discussion of bursters to another paper,
where optically thick solutions will be reported \citep{Wielgus2015}.

The atmospheres presented here can be thought of as an (extreme) example of density {and pressure} inversion. Atmospheric density inversion owing to variations of
opacity has been discussed previously in various contexts.
{In stellar structure theory discussion of density inversion goes back to \cite{1936MNRAS..97..132C}, see also \cite{Sen}. In contemporary papers such layers are often invoked in the context of wind loss\footnote{{In LBV and hypergiant stars ``the most striking propert[y] is the strong density inversion in the outer layers, where a thin gaseous layer floats upon a radiatively supported zone. This is due to the peak in the opacity which forces supra-Eddington luminosities in some layers.'' \citep{1998salg.conf....1C}}}
and stellar pulsations \citep[e.g.,][]{2006MNRAS.368L..57T}.

The possibility of gas pressure inversion was  pointed out by Erika B\"ohm-Vitense, and explained as a result of a change of sign in the effective gravitational acceleration, $g_{\rm eff}$, with $g_{\rm eff}$ being the difference between the true acceleration of gravity and the acceleration owing to radiative forces, which are proportional to the product of opacity and the radiative flux \citep{1958ZA.....46..108B}. This is indeed the cause of the pressure inversion in the atmospheres presented in our paper. 
However, there are several important differences between the results obtained previously and those reported in this work. 

First, the inversion layers familiar from stellar structure theory are caused by variations of opacity,  while} our results are of a different
origin---the (Thomson) opacity in our solutions is uniform, and the density
{(and pressure) inversion arises solely owing to effects of general relativity. More generally, its presence is related to purely geometrical effects that allow the radiative flux to have a different functional dependence on the distance from the source than the acceleration of gravity.

Second, we are presenting optically thin atmospheres, whereas the stellar discussion occurred for optical depths greater than unity. 

Third, in the optically thick case considered by \citet{1973ApJ...181..429J}  density inversion occurs in regions where the luminosity is near-Eddington (although still subcritical), the pressure is dominated by radiation, and the temperature gradient is superadiabatic. While the atmospheres discussed in our paper do require near-Eddington luminosities, none of the other conditions is met. Density inversion occurs only in the region of supercritical luminosity, the radiation pressure is zero, as the radiation freely streams through the optically thin gas, and the temperature gradient is subadiabatic for the isothermal atmosphere, of course, as well as for polytropic atmospheres with $\Gamma<5/3$. Gas-pressure inversion occurs at supercritical luminosities.
The region of pressure inversion coincides with that of density inversion in the optically thin atmospheres presented here.

Fourth, \citet{1973ApJ...181..429J}, in their discussion of optically thick inversion layers in hydrostatic equilibrium, referring to the work of \cite{1970ApJ...159..879L} observe that their own ``results do not alter the well known conclusion that mass-outflow from the surface must result if $L>L^s_{\rm crit}$, where  $L$ and $L^s_{\rm crit}$ are the photospheric values'' of luminosity and its critical value. We note that our atmospheric solutions do ``alter the well known conclusion,'' as the levitating atmospheric shell is in hydrostatic equilibrium even though $L>L_{\rm Edd}$ in (a part of) the  optically thin shell, i.e.,  the luminosity is supercritical above the photosphere of the underlying luminous star. This is possible because \cite{1970ApJ...159..879L} used Newtonian gravity while our solutions are valid in GR.

We conclude with a brief discussion of stability. The atmospheres presented in this paper are convectively stable, as shown by the appropriate Schwarzschild criterion, in the isothermal case and for polytropes with $\Gamma<5/3$. This result can be intuitively understood by realizing that by the law of Archimedes a hot parcel of gas is initially accelerated in the direction opposite to the gradient of pressure, hence the pressure-inversion layer is as stable as the layer of radially decreasing pressure. 
A polytropic atmosphere with $\Gamma=5/3$ is marginally stable to convection. The atmospheres are also Rayleigh-Taylor stable as the denser layers are always ``below" (in the sense of the direction of effective gravity) the less dense layers. Investigation of the normal modes of the atmospheres will be presented elsewhere. Here, we only note that \citep[as shown by][]{Abarca} the radiation drag efficiently damps oscillatory motion in a class of fundamental modes, just as it does the radial and azimuthal motion of test particles \citep{Stahl2012}.
}


%

\section{Acknowledgments}
It is a pleasure to thank Lars Bildsten for a discussion on density inversion.
We thank the anonymous referee for the very helpful comments.
This research was supported in part by the Polish NCN grant UMO-2013/08/A/ST9/00795.

\bibliographystyle{aa}
\bibliography{pmnrathin.bib}

\end{document}